\documentclass[conference]{IEEEtran}
\usepackage{amsmath}
\usepackage{graphicx}
\usepackage{textcomp}
\usepackage{url}
\usepackage{graphicx}
\usepackage[backend=biber, style=numeric-comp, sorting=none]{biblatex}
\usepackage[hidelinks]{hyperref}
\bibliography{ref}

\begin{document}

\title{Design and Analysis of Curved Electrode Configurations for Enhanced Sensitivity in 1-Axis MEMS Accelerometers}

\author{\IEEEauthorblockN{Adhinarayan Naembin Ashok and Adarsh Ganesan}
\IEEEauthorblockA{Department of Electrical and Electronics Engineering,\\Birla Institute of Technology and Science, Pilani - Dubai Campus,\\Dubai International Academic City, Dubai, UAE 345055\\
Email: adarsh@dubai.bits-pilani.ac.in}
}

\maketitle

\begin{abstract}
    This paper presents a comprehensive analytical and simulation-based study of curved electrode geometries for enhancing the sensitivity of MEMS capacitive accelerometers. Expressions for the capacitance between a planar movable electrode and six distinct fixed electrode profiles (biconvex, biconcave, concavo-convex, convexo-concave, plano-convex, and plano-concave) are derived, enabling direct calculation of differential gain and sensitivity as functions of electrode curvature and gap displacement. These analytical models are then rigorously validated using finite element simulations performed using COMSOL Multiphysics under identical bias and boundary conditions. The simulation results demonstrate agreement with the analytical results with a deviation of less than 7\% in all configurations. The results also reveal that biconvex curved electrodes yield the greatest sensitivity improvement over the planar electrodes, with sensitivity monotonically increasing with arc length, while concave and plano-concave designs exhibit reduced performance. The concavo-convex and convexo-concave configurations furthermore introduce polarity inversion in the output voltage, offering additional design flexibility. Importantly, these sensitivity enhancements are achieved without any change in the overall volumetric dimensions of the device or the proof-mass dimensions of the module for achieving higher-resolution inertial sensing.
    
    
\end{abstract}

\section{Introduction}

Miniaturized systems of mechanical and electrical components are known as microelectromechanical systems (MEMS) \cite{najafabadi2024designing}. These systems typically integrate mechanical elements, sensors, actuators, and electronics on a common silicon substrate through microfabrication processes. The typical size of a MEMS device ranges from 20 µm to 1 mm, with the internal mechanical structure often ranging from 1 µm to 100 µm. The rest of the device comprises printed circuitry or is mounted on a printed circuit board (PCB). MEMS technology offers several advantages, including reduced size, low power consumption, mass manufacturability, cost efficiency, and integration with other microelectronic systems \cite{najafabadi2024designing}\cite{li2016novel}. These advantages have positioned MEMS devices as essential components in modern applications such as smartphones, automotive safety systems, industrial monitoring, biomedical implants, and aerospace instrumentation.

Among MEMS devices, accelerometers represent a fundamental type of inertial sensors. Accelerometers measure the acceleration of a body to which they are attached. They can detect both static (e.g., gravity) and dynamic (e.g., movement or vibration) accelerations. A MEMS accelerometer generally consists of a suspended proof mass connected to movable electrodes, placed between fixed electrodes that are anchored to a substrate \cite{li2016novel}. The device operates based on the principle of inertia. When an acceleration is applied, the proof mass experiences a force and displaces relative to the fixed frame. This displacement alters the gap between the moving and fixed electrodes, which in turn changes the capacitance. The variation in capacitance is then converted into a voltage or digital signal using capacitance-to-voltage or capacitance-to-digital converters in the interface circuitry. The capacitive sensing mechanism offers several advantages, such as low noise, low power operation, and high sensitivity to small displacements \cite{babatain2021acceleration}\cite{binali2024comprehensive}. Moreover, capacitive MEMS accelerometers are less affected by temperature fluctuations compared to piezoresistive or piezoelectric counterparts \cite{preeti2019low}. MEMS accelerometers are available in different configurations depending on the axis of sensitivity. Single-axis accelerometers measure acceleration along one direction, while dual-axis and tri-axis accelerometers can sense in two or three orthogonal directions, respectively. Tri-axis MEMS accelerometers have become standard in consumer electronics, enabling orientation sensing, motion tracking, and gesture recognition in smartphones, gaming controllers, and wearable fitness devices \cite{lord2008concurrent}.

The miniaturization enabled by MEMS technology has drastically reduced the size and cost of accelerometers. This has led to their proliferation in numerous domains. In the automotive sector, MEMS accelerometers are used for airbag deployment systems, vehicle stability control, navigation assistance, and crash detection \cite{article}. In consumer electronics, they support features such as screen rotation, step counting, and gaming controls \cite{wu_applications_2021}. In space technology, MEMS accelerometers are vital for performing microgravity experiments, spacecraft attitude determination, and vibration monitoring during launch and re-entry \cite{sabatini2021mems}. Their small size and low weight make them ideal for satellite missions, especially small-scale CubeSats and nanosatellites, where payload constraints are significant. Some space missions also use MEMS accelerometers in inertial navigation units and for scientific measurements in planetary probes \cite{chu2018mems}. In the biomedical field, MEMS accelerometers are gaining importance in health monitoring systems. They can be embedded in wearable devices or smart clothing to monitor body posture, detect falls, and analyse gait patterns. For instance, they have been used to measure the range of motion in patients undergoing rehabilitation and to assess motor impairment in individuals with Parkinson’s disease \cite{lord2008concurrent}\cite{morris2020use}. The ability to continuously monitor physical activity and biomechanical signals makes MEMS accelerometers indispensable in personalized healthcare and telemedicine \cite{tao2021internet}. Furthermore, MEMS accelerometers have significant applications in geophysics and environmental monitoring. In landslide-prone regions, they are embedded in soil or structures to detect subtle shifts and vibrations, which serve as early indicators of slope instability \cite{najafabadi2024designing}\cite{barari2024mems}. Similarly, they play a role in seismic monitoring by detecting tremors and ground motion associated with tectonic activity. This data is valuable for earthquake prediction and for studying the dynamic behaviour of earth materials \cite{ghamari2022mems}.

Recent advances in fabrication techniques and materials have led to further improvements in MEMS accelerometers. Researchers have explored novel designs such as interdigitated comb structures, dual proof-mass systems, and symmetric suspension mechanisms to enhance sensitivity and reduce cross-axis sensitivity \cite{najafabadi2024designing}\cite{hu2022high}. Additionally, the use of high-aspect-ratio deep reactive ion etching (DRIE), SOI wafers, and low-noise readout electronics has pushed the performance of MEMS accelerometers to levels suitable for precision scientific applications. Low-frequency MEMS accelerometers are specifically designed for detecting slow and small motions, such as vibrations of biological tissues or infrastructure health monitoring \cite{preeti2019low}. These accelerometers are particularly sensitive and can detect displacements on the nanometre scale. For example, in structural health monitoring, they are used to detect cracks, corrosion, or fatigue in bridges, dams, and buildings. They help ensure public safety by providing real-time alerts and enabling preventive maintenance strategies \cite{zhu2020mems}. Power management is another critical factor in MEMS accelerometer deployment, particularly in portable and remote systems. Many modern MEMS accelerometers incorporate sleep modes, event-triggered activation, and ultra-low-power signal processing circuits to extend battery life in IoT devices and wearable applications \cite{maluf2020introduction}.

As the demand for autonomous systems grows, MEMS accelerometers are also playing an integral role in robotics, drones, and unmanned vehicles. They contribute to inertial navigation, stabilization, and control algorithms by providing real-time motion data. Combined with gyroscopes and magnetometers, they form inertial measurement units (IMUs), which are the backbone of motion tracking systems used in AR/VR platforms and precision agriculture \cite{tanskanen2023imu}.

This study aims to enhance the sensitivity of a MEMS capacitive accelerometer without altering its overall dimensional constraints, by modifying only the geometry of the fixed electrodes. While conventional designs typically employ planar electrodes, this work investigates alternative curved geometries to improve performance. A comparative analysis is presented between various curved electrode configurations and the standard planar design. Section 2 introduces the proposed accelerometer architecture incorporating curved electrodes. Section 3 details the mathematical modelling and analytical derivation of the device sensitivity. Section 4 provides a comprehensive comparison between the analytical results and COMSOL Multiphysics simulations, validating the effectiveness of the proposed geometries. Section 5 draws out conclusions from the discussions from Sections 2, 3 and 4. 

\section{Description of a 1-Axis MEMS Accelerometer with Curved Electrodes}

Microelectromechanical systems (MEMS) accelerometers transduce inertial forces into electrical signals by exploiting the deflection of a microfabricated proof mass and the resultant change in capacitance between interdigitated electrodes.
A central proof mass is suspended within a cavity by compliant flexures or springs. The mass–spring system has a well‐defined mechanical resonance with mass $m$ and stiffness $k$, which governs its response to acceleration. The proof mass has electrodes attached to it on each side and is free to move as the MEMS module is accelerated, as shown in Fig. 1. Each movable electrode has a pair of fixed electrodes on both sides, called "fixed comb‐drive fingers". These interdigitated electrodes form two capacitors, \( C_1 \) and \( C_2 \), arranged differentially \cite{kovacs1998micromachined}. The device is often encapsulated in a vacuum to reduce damping and improve bandwidth and sensitivity \cite{roukes1991mechanical}. However, for proof of concept, the simulations have been carried out for a device operated in air. \\

When the device is subjected to an acceleration \( a \), an inertial force is generated as \( F = ma \). This force causes a deflection by an amount \( \Delta x \) (Fig. 1). Therefore, \(F = ma = k \Delta x \Rightarrow \Delta x = \frac{ma}{k} \). As the proof mass moves, the gap spacing between movable and fixed comb fingers on one side decreases, while on the opposite side it increases. For comb fingers of same overlapping length, finger thickness and number of finger pairs N, the capacitances $C_1$ and $C_2$ can be expressed as,
\begin{equation*}
C_1 = N C(x - \Delta x) \quad \text{and} \quad C_2 = N C(x + \Delta x)
\end{equation*}
where $C$ is the capacitance between a fixed electrode and a movable electrode. The differential capacitance signal is thus:
\begin{equation*}
\Delta C = C_1 - C_2 = - (C_2 - C_1) = N C(\Delta x)
\end{equation*}
This differential arrangement cancels common‐mode disturbances (e.g., temperature drift, parasitic capacitances) and doubles sensitivity \cite{beeby2004mems}. \\
Now, a differential AC excitation is applied: \( {+V}_{\text{in}} \) to the fixed fingers connected to \( C_2 \), and \( {-V}_{\text{in}} \) to the fixed fingers connected to \( C_1 \). The sum of the incremental charges on \( C_1 \), \( C_2 \), and \( C_{\text{fb}} \) must be zero, where \( C_{\text{fb}} \) is the feedback capacitance.

\begin{equation}
(-C_1 V_{\text{in}}) + (C_2 V_{\text{in}}) + (C_{\text{fb}} V_{\text{out}}) = 0
\end{equation}
\begin{equation}
\Rightarrow V_{\text{out}} = -V_{\text{in}} \frac{C_2 - C_1}{C_1 + C_2}
\end{equation}

\begin{figure}
\centering
\includegraphics[width=7cm,height=6cm]{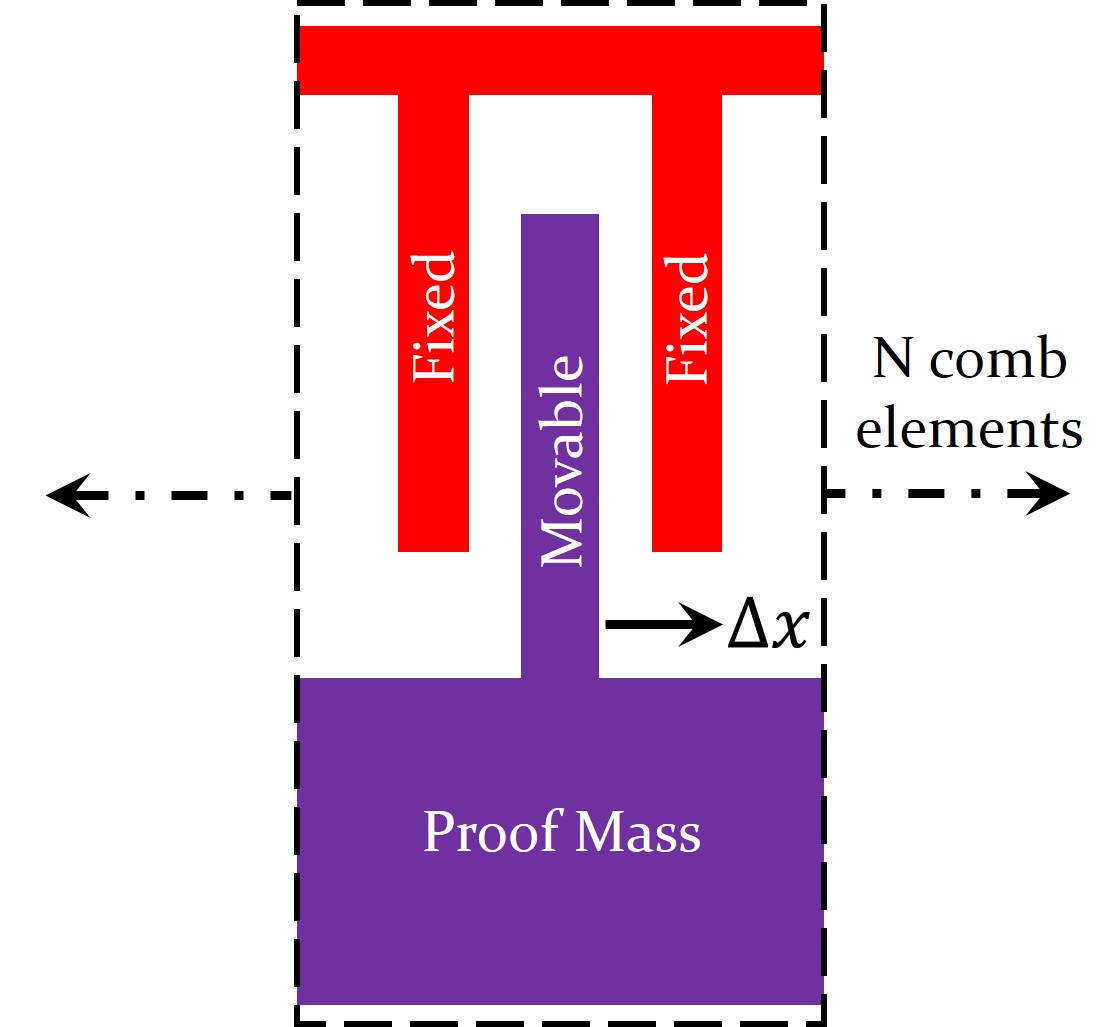}
\caption{Schematic of fixed-movable electrode configuration along with proof mass in MEMS accelerometer module}
\end{figure}

\begin{figure}
\centering
\includegraphics[width=6.6cm,height=7cm]{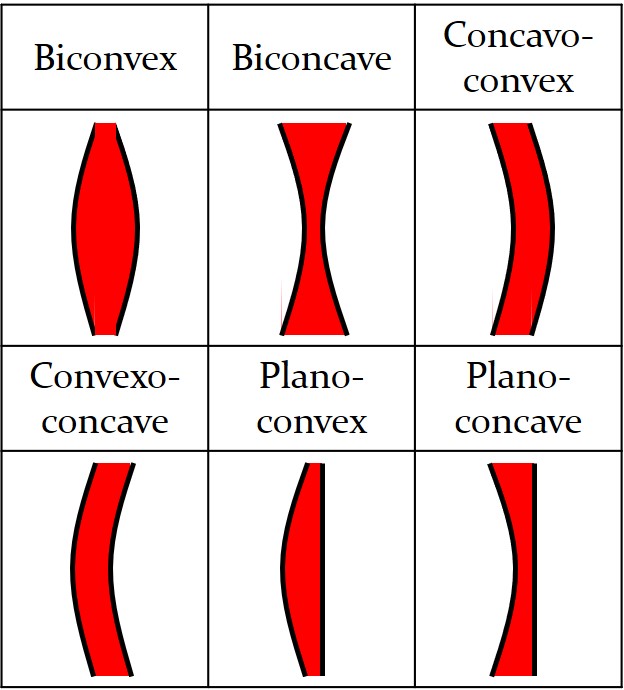}
\caption{Different geometries of electrodes used for the analysis}
\end{figure}

This is the raw gain of the front end \cite{kovacs1998micromachined},\cite{yazdi1998micromachined}.
A common design goal is to have the output span \( \pm V_{\text{in}} \) over the full mechanical range, scaled by the fractional change of the capacitive bridge:
\begin{equation*}
-\frac{C_2 - C_1}{C_1 + C_2} = \pm 1 \quad \text{at full deflection}
\end{equation*}
To exactly achieve Equation (2), we set, 
\begin{equation}
C_{\text{fb}} = C_1 + C_2
\end{equation}
Under this choice, the amplifier automatically compensates for absolute capacitance shifts, minimizing gain drift \cite{roukes1991mechanical}\cite{beeby2004mems}. This also allows the output to span the full excitation voltage, maximizing dynamic range without additional scaling. In practical designs, however, \( C_{\text{fb}} = 2C(\Delta x = 0) = 2C_0 \), which is the nominal capacitance at zero displacement. For our theoretical model and as a proof of concept, we instead set \( C_{\text{fb}} = C_1 + C_2 \) to simplify the analysis and achieve exact normalization under ideal conditions.

Therefore, this proof is fully correct for an ideal, linear, differential capacitive bridge read out by a perfect charge amplifier. In real MEMS accelerometers, one implements trimmed or matched feedback capacitances and compensates for amplifier finite gain and parasitics to approach this ideal performance \cite{roukes1991mechanical}\cite{beeby2004mems}.

The output is demodulated via synchronous detection and low–pass filtered to extract a DC voltage proportional to acceleration \cite{yazdi1998micromachined}. Therefore, a linear dependence exists between the applied acceleration and the resulting displacement of the proof mass. This displacement is sensed through changes in capacitance between the movable and stationary electrodes \cite{comsol2025surface}\cite{langfelder2024comsol}. This data is then used to calibrate sensitivity as, 
\begin{equation}
\text{Sensitivity} = \frac{\Delta V}{\Delta a \ (\text{in } g)}
\end{equation}

The variation between output voltage and acceleration is a linear line with a positive slope. This is observed in almost all MEMS accelerometers. To model the MEMS accelerometer on COMSOL, solid mechanics, electrostatic and electrodynamics are used along with appropriate boundary conditions and mesh controls. Here, the fixed and movable comb electrodes are of same length to simplify the layout design, and maintain linear relationship between the displacements and capacitance which simplifies signal processing and improve measurement accuracy over a fixed range \cite{comsol2025surface}\cite{langfelder2024comsol}\cite{nazdrowicz2017simulink}\cite{divya2015design}.

\begin{table}[h!]
\centering
\caption{Parameters used for simulating different electrode geometries of the MEMS Accelerometer module on COMSOL}
\begin{tabular}{|c|c|}
\hline
Material Used     & Polycrystalline Silicon \\
\hline
Silicon Thickness    & 2 µm\\
\hline
Oxide Thickness    &  1.6 µm\\
\hline
Proof Mass Dimensions & 565 µm $\times$ 100 µm\\
\hline 
Number of self-test fingers & 3 \\
\hline 
Number of electrodes & 21 \\
\hline
Electrode finger width & 4 µm \\
\hline 
Etch hole size & 4 µm $\times$ 4 µm \\
\hline 
Etch hole period & 18 µm \\ 
\hline 
Spring Dimensions & 280 µm $\times$ 2 µm \\ 
\hline 
Spring Gap & 1 µm \\ 
\hline 
Spring Connection width & 4 µm \\ 
\hline 
Anchor Base & 17 µm $\times$ 17 µm \\ 
\hline 
Anchor Radius & 3 µm \\ 
\hline 
Electrode Anchor Radius & 3 µm \\
\hline 
Electrode Length & 120 µm \\ 
\hline 
Electrode Anchor Radius & 3 µm \\ 
\hline 
\end{tabular}
\\
\vspace{1mm}
\begin{center}
\footnotesize
% \textbf{Note:} Load Resistance: $R_L = R_c$  Excitation Acceleration: $a_0 = 0.1\,\mathrm{mg}$ (1g = 9.8 m/s\textsuperscript{2})
\end{center}
\end{table}
% need to add params table here 
% need to add figures here of the things mentioned in para above

% Need to add figure of accelerometer setup here 

Table 1 contains the parameters used in our COMSOL simulations of MEMS accelerometers of different electrode designs. Although parallel plate designs are easy to fabricate and offer robust linearity, their limited capacitance gradient per unit displacement renders them suboptimal when maximum resolution is required at extreme accelerations. Hence, this paper explores the use of curved electrodes in lieu of conventional flat fixed electrodes. Different curved electrode designs considered for the analysis are depicted in Fig. 2.

Curved electrodes offer a significant advancement over traditional flat electrode configurations by introducing a non-uniform gap that varies along the displacement direction \cite{article2}. This variation in the gap profile results in a higher capacitance gradient per unit displacement, which directly translates to enhanced sensitivity and improved signal resolution. Curved geometries allow for more substantial and rapid changes in capacitance as the proof mass moves. This effect becomes particularly valuable when the sensor is required to operate under low acceleration conditions or when maximum resolution is critical.

Choosing the right electrode shape is essential for designing sensitive and precise MEMS accelerometers used in areas like seismic sensing \cite{babatain2021acceleration}, structural monitoring \cite{najafabadi2024designing}, and aerospace \cite{li2016novel}. Flat electrodes offer limited sensitivity due to their constant gap. Curved electrodes, allow better control of the electric field and capacitance gradient. This leads to improved signal resolution. Depending on the application, different electrode curvature configurations can be used to boost sensitivity, while keeping the device dimensions consistent.

In this paper, our choice for curved electrodes is inspired by principles in optics. In optical systems, curved lenses and mirrors are used to focus or direct light more precisely by controlling how it bends across the surface. In a similar way, curved electrodes shape the electric field within a MEMS accelerometer, allowing for more efficient and accurate detection of small movements \cite{article2}. This controlled variation helps improve sensitivity without increasing the size of the device. Just as optical curvature enhances focus and resolution, electrode curvature enhances signal response and measurement precision.

% Need to add figures of electrodes here
\section{Analytical Derivation of Sensitivity for Various Curved Electrode Configurations}

Our chosen curved electrode configurations including biconvex, biconcave, concavo-convex, convexo-concave, plano-convex and plano-concave involve convex/concave surfaces. Hence, to obtain the capacitances between these curved electrodes and the movable planar electrode, we need to derive expressions for the capacitances between the planar and convex/concave surface as a preliminary step.

\subsection{Capacitance between planar surface and convex surface}

Consider the polar coordinate system shown in Fig. 3, where the circular arc, representing the curved electrode, is centred at the origin. The movable electrode, attached to the proof mass, is modelled as an arc segment with an angular extent $\varphi$. Now, to find an expression for the capacitance between planar electrode surface and convex electrode surface, the arc is discretized into infinitesimal angular segments $\theta$. The length of each arc segment is \(Rd\theta\), and the corresponding differential area is \(hRd\theta\), where $h$ denotes the out-of-plane thickness of the electrode.

\begin{figure}
\centering
\includegraphics[width=7.8cm,height=6cm]{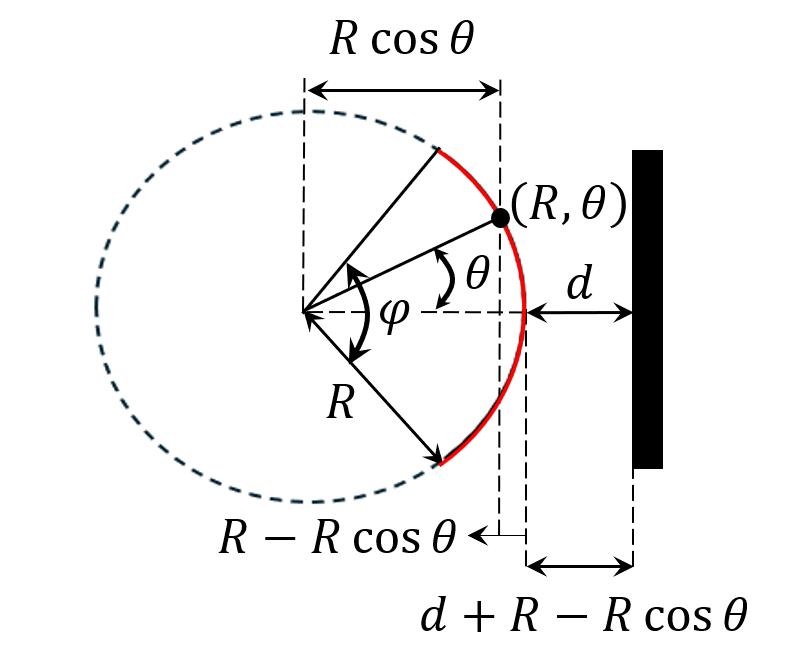}
\caption{Configuration involving a convex surface and a planar surface}
\end{figure}

\begin{figure}
\centering
\includegraphics[width=6.5cm,height=6cm]{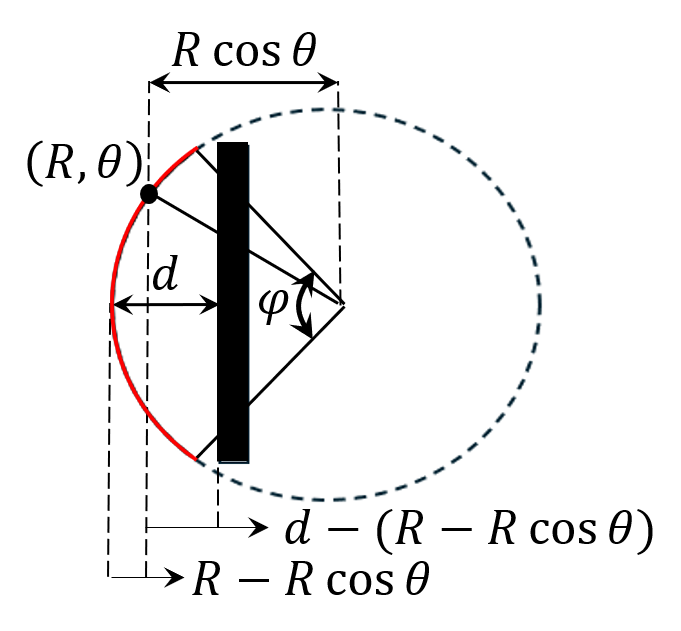}
\caption{Configuration involving a concave surface and a planar surface}
\end{figure}

The capacitance associated with each length element can be written as, 
\begin{equation}
dC = \frac{\varepsilon\,dA}{d(\theta)} = \frac{\varepsilon\,h R\,d\theta}{d+R-R\cos\theta}
\end{equation}

By integrating the expression with the limits \(-\varphi/2\) to \(+\varphi/2\), we get
\begin{equation}
C = \varepsilon h R \int_{-\varphi/2}^{+\varphi/2} \frac{d\theta}{d+R-R\cos\theta}
\end{equation}

The result of this integral is, 
\begin{equation} 
C=\frac{4\varepsilon hR}{\sqrt{d(2R+d)}}\tan^{-1}\left(T\sqrt{\frac{2R+d}{d}}\right)
\end{equation}

where \(T=\tan\left(\frac{\varphi}{4}\right)\)

\subsection{Capacitance between a planar surface and concave surface}

Similarly, to find an expression for the capacitance between a planar electrode surface and a concave electrode surface, \(d(\theta)\) is taken as \( d(\theta) = d + R\cos{\theta}-R\), as shown in Fig.~4. Hence, we have
\begin{equation}
dC=\frac{\varepsilon dA}{d(\theta)}=\frac{\varepsilon hRd\theta}{d+R\cos\theta-R}
\end{equation}
By integrating the expression with the limits \(-\varphi/2\) to \(+\varphi/2\), we get
\begin{equation}
\begin{aligned}
C &= \varepsilon hR\int_{-\frac{\varphi}{2}}^{\frac{\varphi}{2}} \frac{d\theta}{R\cos\theta - R + d} \\
  &= \frac{4\varepsilon hR}{\sqrt{d(2R - d)}}\tanh^{-1}\left(T\sqrt{\frac{2R - d}{d}}\right)
\end{aligned}
\end{equation}

%placed here fig. 5
\begin{figure}
\centering
\includegraphics[width=4.3cm,height=16cm]{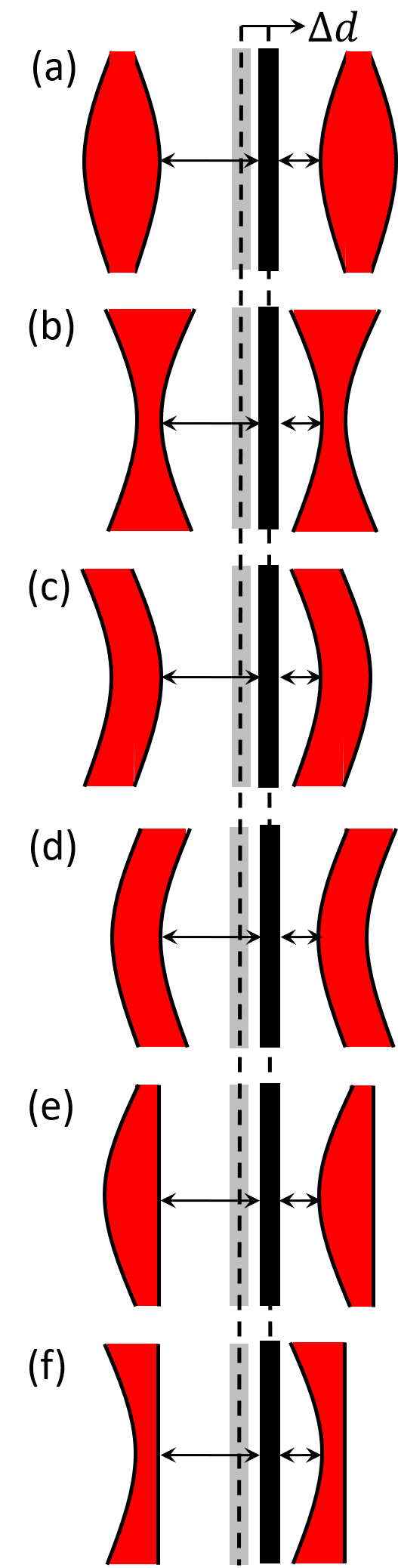}
\caption{(a-f) Displacement of the movable electrode in between the curved electrodes as the MEMS accelerometer module is accelerated for biconvex, biconcave, concavo-convex, convexo-concave, plano-convex and plano-concave configurations respectively}
\end{figure}

\subsection{Capacitance between planar movable electrode and planar fixed electrode}
When both the electrodes are planar, \(d(\theta)\) is merely d. Hence, we have
\begin{equation}
dC=\frac{\varepsilon dA}{d(\theta)}=\frac{\varepsilon hRd\theta}{d}
\end{equation}
By integrating the expression with the limits \(-\varphi/2\) to \(+\varphi/2\), we get
\begin{equation}
C=\varepsilon hR\int_{-\frac{\varphi}{2}}^{\frac{\varphi}{2}}\frac{d\theta}{d}=\frac{\varepsilon hR\varphi}{d}
\end{equation}
Since the arc length \( R\varphi \) is the actual length \( b \) of the planar electrode, we can write  $C$ as \(\frac{\varepsilon h b}{d} \).

\subsection{Differential capacitance for different curved electrode configurations}

A proof mass supports the moving electrode situated between two fixed electrodes. When a balanced excitation voltage is applied, displacement of the mass causes the distance between the movable electrode and one fixed electrode to decrease and the other to increase (Fig. 5). This cause the associated capacitances to change, yielding a differential output voltage that is directly proportional to the applied acceleration. This configuration enhances sensitivity by converting minute mechanical displacements into measurable electrical signals while inherently rejecting common-mode disturbances, thereby improving accuracy and stability in sensing applications \cite{yoon2023highly}. The readout voltage can be calculated as \(V_{out}=-V_{in}\frac{C_2-C_1}{C_2+C_1}\). Since the planar movable electrode gets displaced by \(\Delta d\) due to acceleration as shown in Fig. 5, \(C_1\) becomes \(C_1(d - \Delta d)\) and \(C_2\) becomes \(C_2(d + \Delta d)\).

The formula for voltage of differential capacitance can be used. The equation is as follows,
\begin{equation}
V_{out}=-V_{in}\frac{C_2(d+\Delta d)-C_1(d-\Delta d)}{C_2(d+\Delta d)+C_1(d-\Delta d)}
\end{equation}

From the force equation of an ideal spring, we get \( \Delta d = \frac{m\ddot{x}}{k}=\delta\), where \( k \) is the spring constant, and \( m \) and \( \ddot{x} \) are the total mass and acceleration of the proof mass–movable electrode system, respectively. Hence, the voltage gain \( G(\ddot{x} ) \) can be written as 
\begin{equation}
G\left(\ddot{x} \right)=\frac{V_{out}}{V_{in}}=-\frac{C_2\left(d+\delta\right)-C_1\left(d-\delta\right)}{C_2\left(d+\delta\right)+C_1\left(d-\delta\right)}
\end{equation}

%original for fig 5

\subsubsection{Biconvex}Here, both \( C_1 \) and \( C_2 \) are modelled as planar-convex capacitances, as shown in Fig. 5(a). Hence, we have
\begin{equation}
G(\ddot{x}) = 
\scalebox{0.7}{$
\left(
\frac{
    -\dfrac{
        \tan^{-1}\left(T\sqrt{\dfrac{2R + d + \delta}{d+\delta}}\right)
    }{
        \sqrt{\left(d + \delta\right)\left(2R + d + \delta\right)}
    }
    + \dfrac{
        \tan^{-1}\left(T\sqrt{\dfrac{2R + d - \delta}{d-\delta}}\right)
    }{
       \sqrt{\left(d - \delta\right)\left(2R + d - \delta\right)}
    }
}{
   \dfrac{
        \tan^{-1}\left(T\sqrt{\dfrac{2R + d + \delta}{d+\delta}}\right)
    }{
        \sqrt{\left(d + \delta\right)\left(2R + d + \delta\right)}
    }
    + \dfrac{
        \tan^{-1}\left(T\sqrt{\dfrac{2R + d - \delta}{d-\delta}}\right)
    }{
       \sqrt{\left(d - \delta\right)\left(2R + d - \delta\right)}
    }
}
\right)
$}
\end{equation}
To obtain sensitivity, we differentiate (14) with respect to acceleration. Hence, we get
\begin{equation}
S = \frac{dG(\ddot{x})}{d\ddot{x}} = V_{in} \left( \frac{Z_1(Z_2 + Z_3)}{Z_4} + \frac{Z_5 + Z_6}{\sqrt{Z_4}} \right)
\end{equation}
Here, 
\begin{align*}
Z_1 &= \scalebox{1}{$\left(-\frac{\tan^{-1}(\sqrt{R_+}T)}{\sqrt{D_+}} + \frac{\tan^{-1}(\sqrt{R_-}T)}{\sqrt{D_-}}\right)$}, \\
Z_2 &= -\scalebox{0.9}{$\left(\frac{\sqrt{R_-} M_- \left(-\frac{m}{2k M_-} + \frac{mN_-}{2k M_-^2}\right) T}{\sqrt{D_-}(1 + R_- T^2) R_-} + \frac{\sqrt{R_+} M_+ \left(\frac{m}{2k M_+} - \frac{mN_+}{2k M_+^2}\right) T}{\sqrt{D_+}(1 + R_+ T^2) R_+}\right)$}, \\
Z_3 &= \scalebox{1}{$-\frac{\frac{m(M_- + N_-)}{2k} \tan^{-1}(\sqrt{R_-}T)}{D_-^{3/2}} - \frac{-\frac{m(R_+ + N_+)}{2k} \tan^{-1}(\sqrt{R_+}T)}{D_+^{3/2}}$}, \\
Z_4 &= \scalebox{1}{$\left(\frac{\tan^{-1}(\sqrt{R_+}T)}{\sqrt{D_+}} + \frac{\tan^{-1}(\sqrt{R_-}T)}{\sqrt{D_-}}\right)^2$}, \\
Z_5 &= \scalebox{0.9}{$\left(\frac{\sqrt{R_-} M_- \left(-\frac{m}{2k M_-} + \frac{mN_-}{2k M_-^2}\right) T}{\sqrt{D_-}(1 + R_- T^2) R_-} - \frac{\sqrt{R_+} M_+ \left(\frac{m}{2k M_+} - \frac{mN_+}{2k M_+^2}\right) T}{\sqrt{D_+}(1 + R_+ T^2) R_+}\right)$}, \\
Z_6 &= \scalebox{1}{$\frac{\frac{m(M_- + N_-)}{2k} \tan^{-1}(\sqrt{R_-}T)}{D_-^{3/2}} - \frac{-\frac{m(R_+ + N_+)}{2k} \tan^{-1}(\sqrt{R_+}T)}{D_+^{3/2}}$}, \\
R_\pm &= \scalebox{1}{$\frac{2R + d \pm \delta}{d \pm \delta}$}, \quad
D_\pm = \scalebox{1}{$(d \pm \delta)(2R + d \pm \delta)$}, \\
M_\pm &= \scalebox{1}{$d \pm \delta$}, \quad
N_\pm = \scalebox{1}{$2R + d \pm \delta$}.
\end{align*} 
\subsubsection{Biconcave}
Here, both \( C_1 \) and \( C_2 \) are modelled as planar-concave capacitances, as shown in Fig. 5(b). Hence, we have
\begin{equation}
G(\ddot{x}) =
\scalebox{0.7}{$
\left(
\frac{
    \dfrac{
        \tanh^{-1}\left(
            T
            \sqrt{\dfrac{2R - (d - \delta)}{d - \delta}}
        \right)
    }{
        \sqrt{(d - \delta)(2R - (d - \delta))}
    }
    -
    \dfrac{
        \tanh^{-1}\left(
            T
            \sqrt{\dfrac{2R - (d + \delta)}{d + \delta}}
        \right)
    }{
        \sqrt{(d + \delta)(2R - (d + \delta))}
    }
}{
    \dfrac{
        \tanh^{-1}\left(
            T
            \sqrt{\dfrac{2R - (d - \delta)}{d - \delta}}
        \right)
    }{
        \sqrt{(d - \delta)(2R - (d - \delta))}
    }
    +
    \dfrac{
        \tanh^{-1}\left(T      \sqrt{\dfrac{2R - (d + \delta)}{d + \delta}}
        \right)
    }{
        \sqrt{(d + \delta)(2R - (d + \delta))}
    }
}
\right)
$}
\end{equation}
To obtain sensitivity, we differentiate (16) with respect to acceleration. Hence, we get
\begin{equation}
\begin{aligned}
S(\ddot{x}) &= \frac{dG(\ddot{x})}{d\ddot{x}} \\
&= V_{in} \left( 
\frac{Z_1(Z_3 + Z_4 + Z_5 + Z_6)}{Z_2} 
+ 
\frac{Z_3 - Z_4 - Z_5 + Z_6}{\sqrt{Z_2}} 
\right)
\end{aligned}
\end{equation}
Here, 
\begin{equation*}
\begin{aligned}
Z_1 &= -(C_2 - C_1), \\
Z_2 &= (C_1 + C_2)^2, \\
Z_3 &= \frac{4T \sqrt{\frac{N_-}{D_-}} \left( \frac{m}{2k} + \frac{mN_-}{2kD_-} \right)}{N_- \sqrt{N_- D_-} \left(1 - \frac{N_- T^2}{D_-}\right)}, \\
Z_4 &= \frac{4T \sqrt{\frac{N_+}{D_+}} \left( -\frac{m}{2k} - \frac{mN_+}{2kD_+} \right)}{N_+ \sqrt{N_+ D_+} \left(1 - \frac{N_+ T^2}{D_+}\right)}, \\
Z_5 &= \frac{4\left( \frac{mD_+}{2k} - \frac{mN_+}{2k} \right) \tanh^{-1}\left( T \sqrt{\frac{N_+}{D_+}} \right)}{(N_+ D_+)^{3/2}}, \\
Z_6 &= \frac{4\left( -\frac{mD_-}{2k} + \frac{mN_-}{2k} \right) \tanh^{-1}\left( T \sqrt{\frac{N_-}{D_-}} \right)}{(N_- D_-)^{3/2}}, \\
N_\pm &= 2R - (d \pm \delta), \quad
D_\pm = d \pm \delta.
\end{aligned}
\end{equation*}
\subsubsection{Concavo-convex}
Here, both \( C_1 \) and \( C_2 \) are modelled as planar-convex and planar-concave capacitances respectively, as shown in Fig. 5(c). Hence, we have
\begin{equation}
G(\ddot{x}) = \frac{V_{\text{out}}}{V_{\text{in}}} = 
\scalebox{0.7}{$
\left(
\frac{
    \dfrac{-4 \tanh^{-1} \left(T \sqrt{\dfrac{N_-}{M_+}}\right)}{\sqrt{M_+ N_-}} 
    + 
    \dfrac{4 \tan^{-1} \left(T \sqrt{\dfrac{N_+}{M_-}}\right)}{\sqrt{M_- N_+}}
}{
    \dfrac{4 \tanh^{-1} \left(T \sqrt{\dfrac{N_-}{M_+}}\right)}{\sqrt{M_+ N_-}} 
    + 
    \dfrac{4 \tan^{-1} \left(T \sqrt{\dfrac{N_+}{M_-}}\right)}{\sqrt{M_- N_+}}
}
\right)
$}
\end{equation}
Here, \( M_\pm = d \pm \delta \) and \( N_\pm = 2R \pm d - \delta \).
To obtain sensitivity, we differentiate (18) with respect to acceleration. Hence, we get
\begin{equation}
\begin{aligned}
S &= \frac{dG(\ddot{x})}{d\ddot{x}} \\
  &= 
  \scalebox{0.9}{$
    V_{in} \left( 
      \frac{-\left(-Z_1 + Z_2\right)\left(Z_3 + Z_4 + Z_5 + Z_6\right)}{Z_7^2} 
      + \frac{Z_3 - Z_4 - Z_5 + Z_6}{Z_7} 
    \right)
  $}
\end{aligned}
\end{equation}
Here, 
\begin{align*}
Z_1 &= \frac{4\tanh^{-1}{\left(\sqrt{\frac{N_-}{M_+}}\, T\right)}}{\sqrt{M_+ N_-}}, \\
Z_2 &= \frac{4\tan^{-1}{\left(\sqrt{\frac{N_+}{M_-}}\, T\right)}}{\sqrt{M_- N_+}}, \\
Z_3 &= \frac{4\sqrt{\frac{N_+}{M_-}}\, M_-\left(\frac{-m}{2k M_-}+\frac{m N_+}{2k M_-^2}\right)T}{\sqrt{N_+ M_-}\left(1+\frac{N_+ T^2}{M_-}\right)M_-}, \\
Z_4 &= \frac{4\sqrt{\frac{N_-}{M_+}}\, M_+\left(\frac{-m}{2k M_+}-\frac{m N_-}{2k M_+^2}\right)T}{\sqrt{N_- M_+}\left(1-\frac{N_+ T^2}{M_+}\right)M_+}, \\
Z_5 &= \frac{4\left(\frac{m M_+}{2k}-\frac{m N_-}{2k}\right)\tanh^{-1}{\left(\sqrt{\frac{N_-}{M_+}}\, T\right)}}{\sqrt{M_+ N_-}\left(M_+ N_-\right)}, \\
Z_6 &= \frac{4\left(\frac{m M_-}{2k}+\frac{m N_+}{2k}\right)\tan^{-1}{\left(\sqrt{\frac{N_+}{M_-}}\, T\right)}}{\sqrt{M_- N_+}\left(M_- N_+\right)}, \\
Z_7 &= \left(\frac{4\tanh^{-1}{\left(\sqrt{\frac{N_-}{M_+}}\, T\right)}}{\sqrt{M_+ N_-}}+\frac{4\tan^{-1}{\left(\sqrt{\frac{N_+}{M_-}}\, T\right)}}{\sqrt{M_- N_+}}\right).
\end{align*}

\subsubsection{Convexo-concave}
Here, both \( C_1 \) and \( C_2 \) are modelled as planar-concave and planar-convex capacitances respectively, as shown in Fig. 5(d). Hence, we have
\begin{equation}
\begin{aligned}
G(\ddot{x}) &= \frac{V_{out}}{V_{in}} \\
            &= \scalebox{0.8}{$
            \left(
              \frac{
                \dfrac{4\tanh^{-1}\left(T\sqrt{\dfrac{N_-}{M_-}}\right)}{\sqrt{M_- N_-}} 
                - \dfrac{4\tan^{-1}\left(T\sqrt{\dfrac{N_+}{M_+}}\right)}{\sqrt{M_+ N_+}} 
              }{
                \dfrac{4\tanh^{-1}\left(T\sqrt{\dfrac{N_-}{M_-}}\right)}{\sqrt{M_- N_-}} 
                + \dfrac{4\tan^{-1}\left(T\sqrt{\dfrac{N_+}{M_+}}\right)}{\sqrt{M_+ N_+}} 
              }
            \right)
            $}
\end{aligned}
\end{equation}
Here, 
\begin{equation*}
M_\pm = \scalebox{1}{$d \pm \delta$}, \quad N_\pm = \scalebox{0.7}{$2R \pm d + \delta$}
\end{equation*}
To obtain sensitivity, we differentiate (20) with respect to acceleration. Hence, we get
\begin{equation}
\begin{aligned}
S &= \frac{dG(\ddot{x})}{d\ddot{x}} \\
  &= \scalebox{1}{$V_{in} \left( 
    \frac{-\left(-Z_1 + Z_2\right)\left(Z_3 + Z_4 + Z_5 + Z_6\right)}{Z_7^2} 
    + \frac{Z_3 - Z_4 - Z_5 + Z_6}{Z_7} 
  \right)$}
\end{aligned}
\end{equation}
Here, 
\begin{align*}
Z_1 &= \frac{4\tanh^{-1}\left(\sqrt{\frac{N_-}{M_-}}\, T\right)}{\sqrt{M_- N_-}}, \\
Z_2 &= \frac{4\tan^{-1}\left(\sqrt{\frac{N_+}{M_+}}\, T\right)}{\sqrt{M_+ N_+}}, \\
Z_3 &= \frac{4\sqrt{\frac{N_+}{M_+}}\, M_+ \left( \frac{-m}{2k M_+} + \frac{m N_+}{2k M_+^2} \right) T}
           {\sqrt{N_+ M_+}\left(1 + \frac{N_+ T^2}{M_+}\right) M_+}, \\
Z_4 &= \frac{4\sqrt{\frac{N_-}{M_-}}\, M_- \left( \frac{-m}{2k M_-} - \frac{m N_-}{2k M_-^2} \right) T}
           {\sqrt{N_- M_-}\left(1 - \frac{N_+ T^2}{M_-}\right) M_-}, \\
Z_5 &= \frac{4\left( \frac{m M_-}{2k} - \frac{m N_-}{2k} \right) \tanh^{-1}\left( \sqrt{\frac{N_-}{M_-}}\, T \right)}
           {\sqrt{M_- N_-}\left(M_- N_-\right)}, \\
Z_6 &= \frac{4\left( \frac{m M_+}{2k} + \frac{m N_+}{2k} \right) \tan^{-1}\left( \sqrt{\frac{N_+}{M_+}}\, T \right)}
           {\sqrt{M_- N_+}\left(M_- N_+\right)}, \\
Z_7 &= \left( 
           \frac{4\tanh^{-1}\left( \sqrt{\frac{N_-}{M_-}}\, T \right)}{\sqrt{M_- N_-}} 
           + \frac{4\tan^{-1}\left( \sqrt{\frac{N_+}{M_+}}\, T \right)}{\sqrt{M_+ N_+}} 
       \right).
\end{align*}

\subsubsection{Plano-convex}
Here, \( C_1 \) and \( C_2 \) are modelled as planar-convex and planar-planar capacitances respectively, as shown in Fig. 5(e). Hence, we have
\begin{equation}
G(\ddot{x}) = \frac{V_{\text{out}}}{V_{\text{in}}} = 
\scalebox{0.6}{$
\left(
\frac{
    \dfrac{4\varepsilon h R}{\sqrt{(d - \delta)(2R + (d - \delta))}} 
    \tan^{-1} \left( T \sqrt{ \dfrac{2R + (d - \delta)}{d - \delta} } \right) 
    - \dfrac{\varepsilon b h}{d + \delta}
}{
    \dfrac{4\varepsilon h R}{\sqrt{(d - \delta)(2R + (d - \delta))}} 
    \tan^{-1} \left( T \sqrt{ \dfrac{2R + (d - \delta)}{d - \delta} } \right) 
    + \dfrac{\varepsilon b h}{d + \delta}
}
\right)
$}
\end{equation}
To obtain sensitivity, we differentiate (22) with respect to acceleration. Hence, we get
\begin{equation}
S(\ddot{x}) = \frac{dG(\ddot{x})}{d\ddot{x}} = 
\scalebox{1}{$
V_{in} \left( 
\frac{Z_1 \left( Z_2 + Z_3 - Z_4 \right)}{Z_5} 
+ \frac{Z_2 + Z_3 + Z_4}{\sqrt{Z_5}} 
\right)
$}
\end{equation}
Here, 
\begin{align*}
Z_1 &= -(C_2 - C_1), \\
Z_2 &= \frac{4Rt \sqrt{\frac{N}{D}} \left( -\frac{m}{2k} + \frac{mN}{2kD} \right) T}{N \sqrt{ND} \left( 1 + \frac{NT^2}{D} \right)}, \\
Z_3 &= \frac{4Rt \left( \frac{mD}{2k} + \frac{mN}{2k} \right) \tan^{-1} \left( T \sqrt{\frac{N}{D}} \right)}{(ND)^{3/2}}, \\
Z_4 &= \frac{bmt}{k(d + \delta)^2}, \\
Z_5 &= (C_1 + C_2)^2, \\
N &= 2R + d - \delta, \quad D = d - \delta.
\end{align*}

\subsubsection{Plano-concave}
Here, \( C_1 \) and \( C_2 \) are modelled as planar-concave and planar-planar capacitances respectively, as shown in Fig. 5(f). Hence, we have
\begin{equation}
\scalebox{0.7}{$
\begin{aligned}
G(\ddot{x}) &= \frac{V_{\text{Out}}}{V_{\text{in}}} \\
&= \left(
\frac{
    \dfrac{4\varepsilon hR}{\sqrt{(d - \delta)(2R - (d - \delta))}} 
    \tanh^{-1} \left( 
        T \sqrt{\dfrac{2R - (d - \delta)}{(d - \delta)}} 
    \right) 
    - \dfrac{\varepsilon b h}{d + \delta}
}{
    \dfrac{4\varepsilon hR}{\sqrt{(d - \delta)(2R - (d - \delta))}} 
    \tanh^{-1} \left( 
        T \sqrt{\dfrac{2R - (d - \delta)}{(d - \delta)}} 
    \right) 
    + \dfrac{\varepsilon b h}{d + \delta}
}
\right)
\end{aligned}
$}
\end{equation}
To obtain sensitivity, we differentiate (24) with respect to acceleration. Hence, we get
\begin{equation}
\begin{aligned}
S(\ddot{x}) &= \frac{dG(\ddot{x})}{d\ddot{x}} \\
            &= V_{in} \scalebox{0.75}{$\left( 
                \frac{Z_1\left(Z_2 + Z_3 - Z_4\right)}{Z_5} 
                + \frac{Z_2 + Z_3 + Z_4}{\sqrt{Z_5}} 
            \right)$}
\end{aligned}
\end{equation}
Here, 
\begin{align*}
Z_1 &= -(C_2 - C_1), \\
Z_2 &= \frac{4Rt\sqrt{\frac{N}{D}}\left(-\frac{m}{2k}+\frac{mN}{2kD}\right)T}{N\sqrt{ND}\left(1+ \frac{NT^2}{D}\right)}, \\
Z_3 &= \frac{4Rt\left(\frac{mD}{2k}+\frac{mN}{2k}\right)\tanh^{-1}\left(T\sqrt{\frac{N}{D}}\right)}{(ND)^{3/2}}, \\
Z_4 &= \frac{bmt}{k(d+\delta)^2}, \\
Z_5 &= (C_1 + C_2)^2, \\
N &= 2R - (d - \delta), \quad
D = d - \delta.
\end{align*}
The sensitivity obtained in the subsections 1-6 corresponds to one comb comprising one movable electrode and two fixed electrodes. Since there are $N$ such combs, the net sensitivity ($S_{\text{net}}$) offered by the MEMS accelerometer will be $S_{\text{net}} = nS$.

\section{Verification and Validation Using Finite Element Modeling}

The MEMS accelerometer is modelled and simulated in COMSOL Multiphysics, leveraging its finite‐element–based multiphysics environment to capture coupled mechanical and electrical phenomena. Specifically, the solid mechanics interface resolves the device’s structural deformation and dynamic response under applied acceleration, while the electrostatics interface computes the evolving electric field distribution and corresponding capacitance changes between electrodes where transient electromagnetic effects are relevant. The electrodynamics interface is also employed to account for time‐dependent interactions. \\ 

\subsection{Validating the analytical results with simulation results of different electrodes}

\begin{figure*}
\centering
\includegraphics[width=16cm,height=8.5cm]{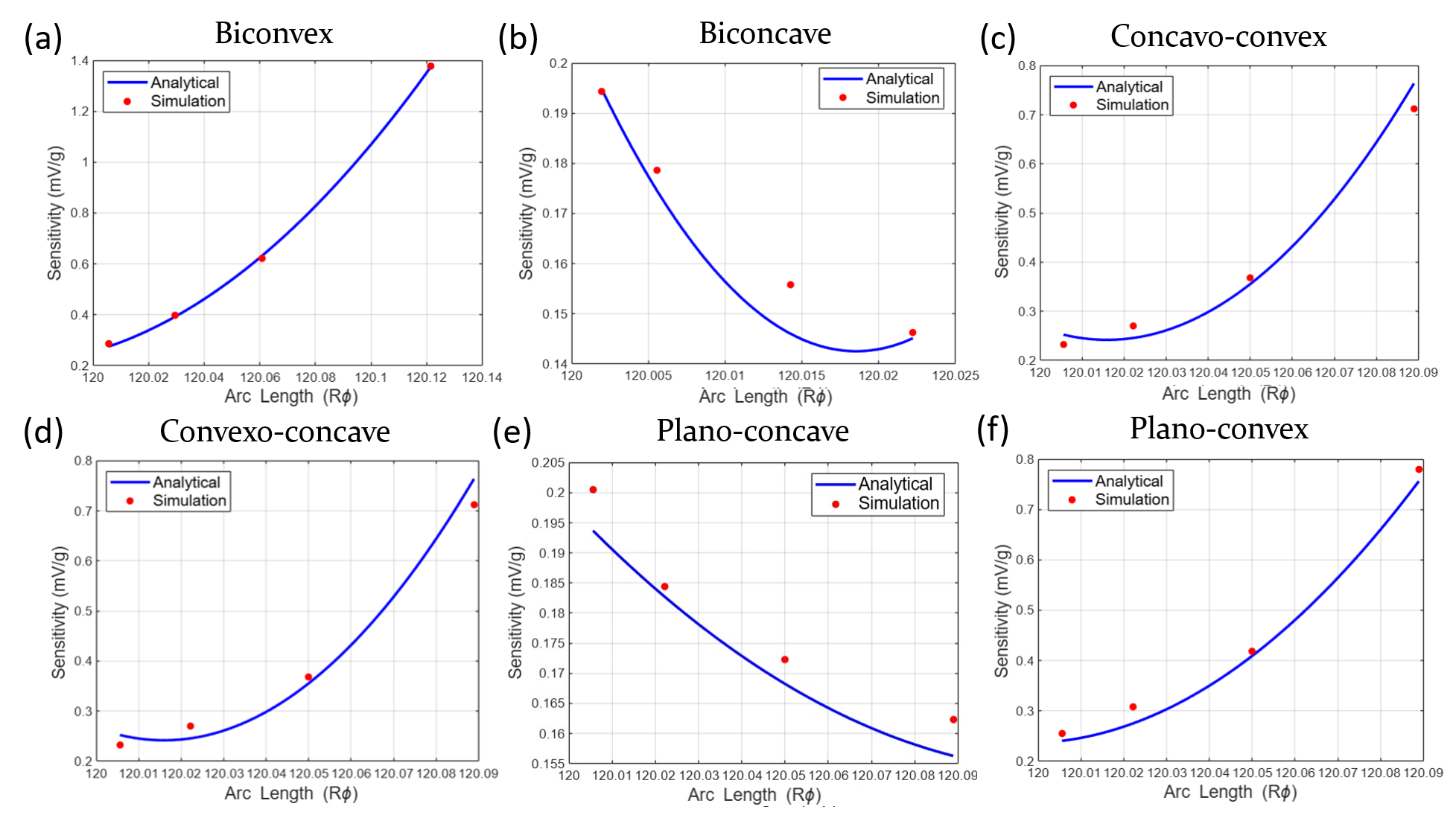}
\caption{(a-f) Comparison of analytical and simulation results of sensitivity vs. arc length for biconvex, bi-
concave, concavo-convex, convexo-concave, plano-convex and plano-concave
configurations respectively}
\end{figure*}

\begin{figure}
\centering
\includegraphics[width=8cm,height=12cm]{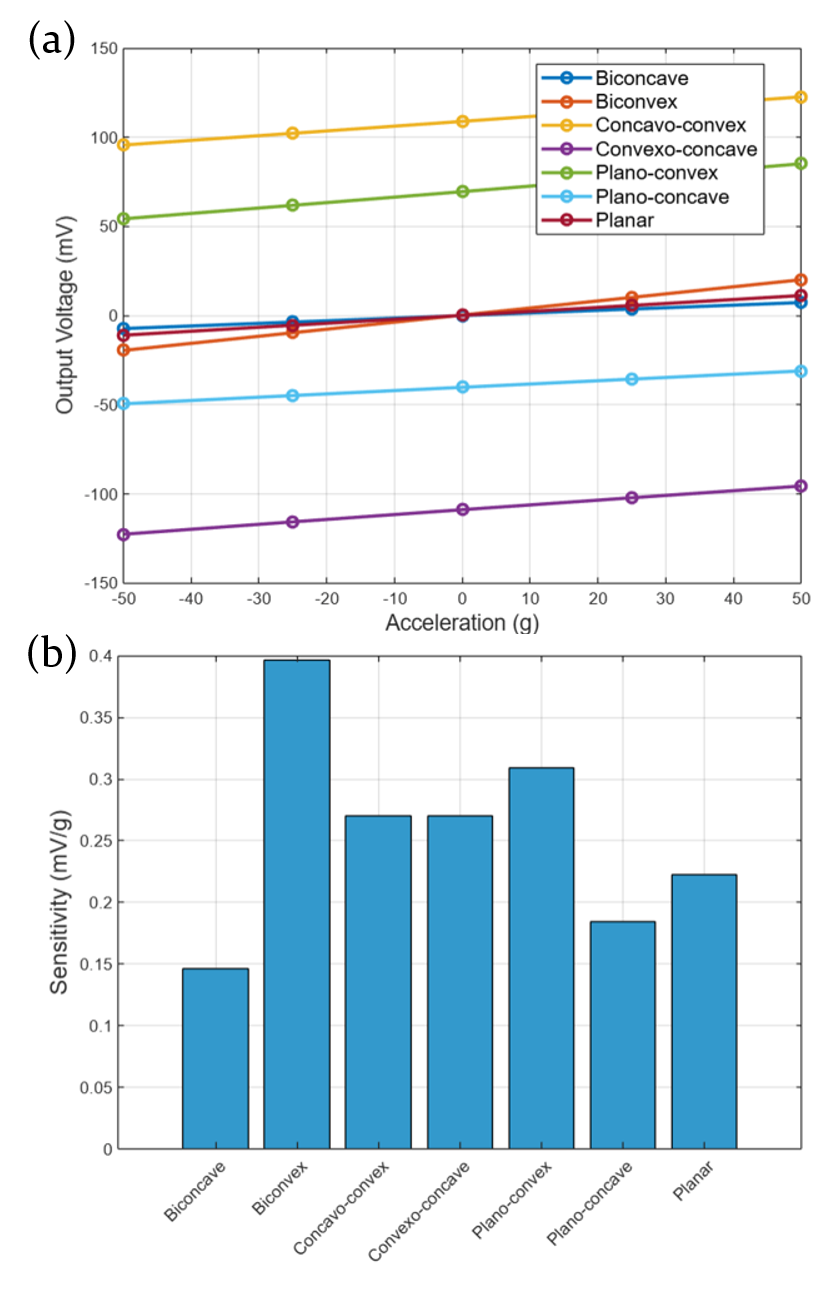}
\caption{(a-b) Output voltage vs acceleration and sensitivity for different electrode configurations respectively}
\end{figure}

To validate the analytical model using COMSOL simulation, the governing equations are implemented in a Python environment (Python 3.10) using NumPy and SciPy for numerical evaluation. For each electrode geometry, the symbolic expressions for capacitance change and differential voltage output are discretized under the same bias conditions as carried out in the FEM study. To facilitate direct visual comparison, both the analytical and FEM derived responses are plotted in Fig. 6. 

% Table and figure for biconvex
In biconvex (Fig. 6(a)) and plano-convex electrode configurations (Fig. 6(f)), the curvature directs the fixed electrodes outward towards the movable element. Hence, the gap between the electrodes decreases with an increase in arc length. This inward displacement of the fixed electrode’s midpoint enhances the baseline capacitance and amplifies the change in capacitance for a given proof‐mass displacement. As a result, the electrostatic coupling strengthens with longer arcs, yielding an increasing sensitivity trend. 

% Table and figure for biconcave
However, in biconcave (Fig. 6(b)) and plano-concave (Fig. 6(e)) electrode configurations, an increased arc length causes reduction in sensitivity because the inward curvature enlarges the central separation between the fixed and movable electrodes. This geometric effect directly weakens the electrostatic transduction mechanism, so that, despite the greater overall electrode surface, the capacitance variation per unit motion is reduced. \

The results of concavo-convex and convexo-concave are depicted in Fig. 6(c) and Fig. 6(d), respectively. \
% Table and figure for convexo-concave and its counterpart
It must be noted that in concavo-convex versus convexo-concave geometries, the sign of the differential capacitance changes and hence the polarity of the output voltage reverses because the moving electrode always sees a gap change of opposite sign on the two sides of its neutral position. Thus, simply by swapping the direction in which the fixed electrodes bow, one inverts the sign of $\Delta C$ for a given displacement, and the sensor’s output voltage polarity flips accordingly. Therefore, concavo-convex and convexo-concave have similar sensitivity dependences with arc length. \

The simulation results thus validate our mathematical model of acceleration sensitivity and henceforth align with the physical characteristics introduced by each curvature configuration.

\subsection{Sensitivity Comparison across different geometries of the electrodes}

A systematic sensitivity analysis defined as the change in output signal (voltage or capacitance) per unit acceleration is carried out across multiple electrode geometries. Each configuration is subjected to identical boundary conditions and excitation profiles to ensure a fair comparison. Fig. 7(a) compares the output voltage vs acceleration plots of different electrode geometries, and Fig. 7(b) compares the slopes of these plots i.e. sensitivity (mV/g). These plots provide a quantitative basis for selecting and optimizing electrode geometries in future MEMS accelerometer designs. 
% table and figures 

\section{Conclusion}
A strategic modification of fixed electrode geometry influences capacitive transduction and overall sensitivity of MEMS accelerometers. Among the six curved electrode profiles examined, the biconvex configuration stands out, delivering the highest sensitivity while maintaining linearity and common mode rejection. The analytical predictions and COMSOL simulations match well with deviations being less than 7\%. This validates the underlying modelling approach and confirms the practical viability of curved electrode designs. Biconcave and plano-concave geometries, by contrast, demonstrate reduced sensitivity due to increased midpoint gap spacing, highlighting the critical role of curvature direction. The polarity reversal observed between concavo-convex and convexo-concave geometries further expands the design space for differential readout schemes, offering designers additional flexibility in tailoring device response for specific applications. Beyond sensitivity enhancement, curved electrode architectures introduce new avenues for optimizing bandwidth and noise performance. By judiciously selecting the curvature parameters and integrating them with established proof mass and suspension designs, it becomes possible to fine-tune the trade-off between mechanical stiffness and electrical transduction. 

Future work can focus on fabrication-aware simulation workflows that replicate key MEMS processing steps like photolithographic patterning, sacrificial layer etching, structural release, and circuitry integration to quantify how variability in each stage alters electrode profiles and device parameters. Integrating these fabrication simulations into a unified design-to-fabrication pipeline will enable accurate performance forecasts from process settings alone, eliminating the need for initial physical prototypes. Ultimately, the incorporation of curved electrodes presents a low complexity, footprint preserving strategy to push the performance envelope of MEMS accelerometers. High-sensitive curved electrode configurations can also offer engineering benefits in the development of emerging nonlinear MEMS devices including accelerometers \cite{agrawal2013observation}\cite{thiruvenkatanathan2011limits}\cite{zhao2016review}\cite{ganesan2017phononic}\cite{zhang2025coherent}.

\section*{Acknowledgments}
The authors sincerely thank School of Interdisciplinary Research
and Entrepreneurship (SIRE), BITS Pilani, for providing the
funding support for this project through its SPARKLE program. The authors thank Taha Ashraf Ali Shaikh and Jai Aadhithya Ramesh for helping us in a few calculations.

\printbibliography

\end{document}